\documentclass[aip,%
               jcp,%
               preprint,%
               citeautoscript,%
               bibnotes,%
               amsfonts,%
               amssymb,%
               amsmath,%
               a4paper,%
               titlepage,%
               fleqn,%
               endfloats*%
               ]{revtex4-1}

\usepackage{graphicx}
\usepackage{xspace}
\usepackage{float}
\usepackage{color}

\begin{document}

\newcommand{\hoch}[1]{$^{\text{#1}}$}
\newcommand{\tief}[1]{$_{\text{#1}}$}
\def\MP{\text{MP2}\xspace}
\def\gm{\mu}
\def\gn{\nu}
\def\gk{\kappa}
\def\gl{\lambda}
\def\po{\underline{P}}
\def\pv{\bar{P}}
\def\ka{\tilde{a}}
\def\kb{\tilde{b}}
\def\ki{\tilde{i}}
\def\kj{\tilde{j}}
\def\re#1{\text{Re}\left(#1\right)}
\def\im#1{\text{Im}\left(#1\right)}

\title{Laplace-transformed atomic orbital-based M{\o}ller--Plesset perturbation theory
       for relativistic two-component Hamiltonians}

\author{Benjamin Helmich-Paris}
\email{b.helmichparis@vu.nl}
\affiliation{Section of Theoretical Chemistry, VU University Amsterdam, 
             De Boelelaan 1083, 1081 HV Amsterdam, The Netherlands}

\author{Michal Repisky}
\email{michal.repisky@uit.no}
\affiliation{CTCC, Department of Chemistry, UiT The Arctic University of Norway, 
             N-9037 Trom{\o} Norway}

\author{Lucas Visscher}
\email{l.visscher@vu.nl}
\affiliation{Section of Theoretical Chemistry, VU University Amsterdam, 
             De Boelelaan 1083, 1081 HV Amsterdam, The Netherlands}

\date{\today}

\begin{abstract}
We present a formulation of Laplace-transformed atomic orbital-based
second-order M{\o}ller--Plesset perturbation theory (MP2) energies
for two-component Hamiltonians in the Kramers-restricted formalism.
This low-order scaling technique can be used to enable correlated relativistic 
calculations for large molecular systems.
We show that the working equations to compute the relativistic MP2 energy 
differ by merely a change of algebra (quaternion instead of real) from their non-relativistic counterparts.
With a proof-of-principle implementation we study the effect
of the nuclear charge on the magnitude of half-transformed integrals
and show that for light elements spin-free and spin-orbit MP2 energies
are almost identical.
Furthermore, we investigate the effect of separation of charge distributions
on the Coulomb and exchange energy contributions,
which show the same long-range decay with the inter-electronic / atomic
distance as for non-relativistic MP2.
A linearly scaling implementation is possible if the proper
distance behavior is introduced to the quaternion
Schwarz-type estimates as for non-relativistic MP2.
\end{abstract}

\maketitle

\section{Introduction}
The most accurate way amongst today's standard approaches 
to account for relativistic effects 
in molecules is to solve the Dirac equation for the large (LC) 
and small two-component (SC) part of the wave function.\cite{Dyall2007}
Employing the full four-component (4C) methodology of the Dirac equation 
yields solutions for both positronic and electronic states.
An accurate approximation to the Dirac equation 
that gives electronic --- positive energy --- solutions only is the 
exact (X) two-component (2C) method,\cite{Ilias2005,*Kutzelnigg2005,*Kutzelnigg2006,*Ilias2007}
in which the Dirac Hamiltonian is reduced to a 2 x 2 matrix form.
While exact for a given one-body operator, the X2C method usually includes the approximation of
omitting the picture-change transformation of two-electron integrals.
Especially for correlated calculations, this makes the algorithm very efficient as molecular integrals 
then require evaluation of only LC atomic-orbital (AO) two-electron integrals.
Since the SC AO basis can be up to 3~times larger than the LC AO basis,
if the unrestricted kinetic balance condition is employed,
the overhead of the integral calculation and transformation 
is then reduced immensely.\cite{Peng2007,*Liu2009,*Sikkema2009}

With a relativistic 2C Hamiltonian the overhead of
the Dirac equation is thus reduced, but accurate wave function methods
still feature a steep scaling of the computational work with the 
system size $N$, e.g.\ 
second-order M{\o}ller--Plesset perturbation theory\cite{Moeller1934} (MP2) 
scales with $\mathcal{O}(N^5)$.
To overcome this computational bottleneck, several reduced scaling
approaches were proposed for MP2 over the years.\cite{Haeser1992,Haeser1993,Ayala1999,Doser2009,Maurer2014,*Hetzer2000,*Jung2004,*Yang2011,*Schmitz2013}
One of the most successful approaches originates from
Alml{\"o}f\cite{Almloef1991,Haeser1992} and H{\"a}ser\cite{Haeser1993}
where the MP2 correlation energy
is formulated purely in the AO basis by means of 
a numerical Laplace transformation (LT)
of orbital-energy denominators. 
By employing estimates for integral screening,
low-order scaling implementations
were proposed by H{\"a}ser\cite{Haeser1993} and Ayala and Scuseria.\cite{Ayala1999}
Ochsenfeld and his co-workers showed how to achieve linear scaling
of all computationally demanding steps by accounting for the distance-dependent decay of
integrals,\cite{Lambrecht2005a,Doser2009,Maurer2013}
and their implementation currently allows for calculations on molecules with 1000~atoms and more on a
single core.\cite{Maurer2013}
The same authors have also presented a promising computational performance
for large AO basis sets that include high angular momentum and / or diffuse basis functions
when the LT AO-based approach is combined with Cholesky
decomposition of two-electron integrals.\cite{Maurer2014}
Alternatively, the performance of the LT AO-based approach when large basis sets are employed 
can also be remedied with a modern formulation of explicitly correlated MP2-F12\cite{Klopper1987,*Kutzelnigg1991,*Klopper2006a,Li2012,*Tenno2012}
in the AO basis, which became available recently.\cite{Hollman2013}

In the current work, 
we present for the first time formulae for LT AO-based MP2 correlation energy 
for relativistic 2C Hamiltonians in a Kramers-restricted (KR) formalism.
The additional spin orbit (SO)-induced complexity of the formulae is discussed by comparing
with non-relativistic (NR) LT AO-MP2 and conventional KR spinor-based approaches.
Furthermore, we present an adaptation of the Schwarz-type estimates, 
which are employed to screen shell quadruple contributions to the MP2 correlation energy,
to the 2C formalism.
Those Schwarz-type estimates\cite{Haeser1993} pave the way to develop a linearly scaling 
2C LT AO-MP2 implementation once they are combined with the proper
distance-dependence terms.\cite{Maurer2013}
With a proof-of-principle implementation we analyze the
effect of the nuclear charge and the separation of electronic charge distributions
on the correlation energy contributions.

\section{Theory} \label{theory}

\subsection{MP2 energy for 1C, 2C, and 4C Hamiltonians}

The MP2 correlation energy is given by 
\begin{align} \label{emp2}
E_{\MP} 
&= - \frac{1}{4} \sum_{IJAB} \frac{1}{\Delta_{AIBJ}} | (IA||JB) |^2  \\
&= - \frac{1}{2} \sum_{IJAB} \frac{1}{\Delta_{AIBJ}} 
 \left( (BJ|AI) - (AJ|BI) \right) (IA|JB) \label{emp2b} \\
&= E_J - E_K
\text{.}
\end{align}
In Eq.\ \eqref{emp2}, $(IA||JB)$ are anti-symmetrized two-electron integrals $(IA|JB)$
that involve active occupied ($I,J$) and
active virtual ($A,B$) spinors;
$\Delta_{AIBJ}$ is the denominator 
\begin{align}\label{denom}
\Delta_{AIBJ} = \varepsilon_A-\varepsilon_I+\varepsilon_B-\varepsilon_J
\end{align}
that comprises real orbital energies  $\varepsilon_P$ with $P=I,A,J,\text{ and }B$.
Note that Eq.\ \eqref{emp2} is valid for both NR and relativistic 
Hamiltonians when spin orbitals and 4C (or 2C) spinors are
employed, respectively. This equation furthermore only assumes a single 
determinant reference wave function and is therefore applicable to both restricted and
unrestricted Hartree--Fock wave functions. In practice, one usually deals with closed-shell 
systems and it is therefore efficient to make use of Kramers symmetry\cite{Kramers1930}
in the Hartree--Fock self-consistent field procedure. Spinor energies are then doubly degenerate and 
the number of unique variational parameters is reduced by a factor of two as the expansion 
coefficients for pairs of spinors are related. We will assume such a restricted optimization
and label two spinors that form a Kramers pair by the lowercase symbols $p$ and $\tilde{p}$.
These spinors can be expressed in a set of real AO basis functions $\chi_\mu$ as
\begin{align}\label{LCAO}
 \phi_p &= 
 \begin{pmatrix}
  \phi_p^{\alpha}   \\
  \phi_p^{\beta}
 \end{pmatrix}
 =
 \begin{pmatrix}
    \sum_{\mu} \chi_{\mu} C_{\mu p}^{\alpha} \\
    \sum_{\mu} \chi_{\mu} C_{\mu p}^{\beta}
 \end{pmatrix}
 &
  \phi_{\tilde{p}} &= 
 \begin{pmatrix}
  \phi_{\tilde{p}}^{\alpha}  \\
  \phi_{\tilde{p}}^{\beta}
 \end{pmatrix}
=
  \begin{pmatrix}
    \sum_{\mu} \chi_{\mu} (-C_{\mu p}^{\beta *} ) \\
    \sum_{\mu} \chi_{\mu} ( C_{\mu p}^{\alpha *}) 
 \end{pmatrix}
\text{;}
\end{align}\\
a form which clearly displays the Kramers relation between the coefficients.
This expansion is valid for NR, scalar-relativistic (SR), and SO relativistic spinor optimization and can
be defined to encompass both 2C and 4C spinors. In the latter case, the basis set is
divided into separate sets of LC and SC expansion functions and coefficients,
which describe respectively the upper and lower components of the 4C spinor.

In the X2C approximation, that we focus on in the current work, the Dirac--Coulomb Hamiltonian
is reduced to an effective 2C form and picture change corrections on the two-electron operator are neglected. 
With these approximations AO integrals reduce to their familiar NR form
\begin{align}
(\kappa\lambda|\mu\nu)=\int \int \chi_{\kappa}(\mathbf{r}_1) \chi_{\lambda}(\mathbf{r}_1) \frac{1}{r_{12}} \chi_{\mu}(\mathbf{r}_2) \chi_{\nu}(\mathbf{r}_2) d\mathbf{r}_1 d\mathbf{r}_2
\end{align}
and can be evaluated using standard techniques.

Relativity is manifest only in the MO coefficients, with scalar relativistic effects merely changing their value but not changing
the structure of the coefficient matrices. NR and SR
approaches have real, block diagonal, coefficient matrices 
$\mathbf{C}_p^{\alpha}=\mathbf{C}_{\tilde{p}}^{\beta}; \mathbf{C}_p^{\beta}=\mathbf{C}_{\tilde{p}}^{\alpha}=\mathbf{0}$, 
while SO approaches yield complex matrices for which $\mathbf{C}_p^{\beta}\neq 0$. It is the latter effect 
which makes relativistic correlated calculation more expensive than their NR counterparts.

\subsection{LT AO-based MP2 for 1C Hamiltonians}
By means of the (numerical) LT,
it has been shown initially by Alml{\"o}f and H{\"a}ser\cite{Almloef1991,Haeser1992,Haeser1993} that
an expansion of the orbital-energy denominator (Eq. \eqref{denom}) according to
\begin{align} \label{laplace}
\frac{1}{\Delta_{aibj}} &= \int_0^{\infty} \exp(-\Delta_{aibj}\, t) dt
 \approx \sum_{z=1}^{n_z} \omega_{z}\, \exp(-\Delta_{aibj}\, t_{z})
\end{align}
results in a formulation for the Coulomb $(J)$ and 
exchange contribution $(K)$,\cite{Haeser1993,Ayala1999,Lambrecht2005a,Surjan2005}
\begin{align}\label{emp2lap}
 & E_{\MP} = \sum_{z=1}^{n_z} e_J^{(z)}  - e_K^{(z)} 
 \text{,}
\end{align}
solely in terms of intermediates in the AO basis. 
In Eqs.\ \eqref{laplace} -- \eqref{emp2lap} $n_z$ denotes a pre-defined number of quadrature points
represented by $\{\omega_{z},t_{z}\}$.
For the NR Hamiltonian and restricted HF reference wave functions,
occupied
\begin{align} \label{psdeno_nr}
\mathbf{\po}^{(z)} 
 &= |\omega_z|^{1/4}\, \mathbf{C}^o \, \exp(+\mathbf{\boldsymbol\varepsilon}^o t_z)\, (\mathbf{C}^o)^{T}  \notag \\
 &= |\omega_z|^{1/4}\, \exp(+t_z\, \mathbf{P}\, \mathbf{F})\, \mathbf{P}
\end{align}
and virtual pseudo-density matrices
\begin{align} \label{psdenv_nr}
\mathbf{\pv}^{(z)}  
 &= |\omega_z|^{1/4}\, \mathbf{C}^v\, \exp(-\mathbf{\boldsymbol\varepsilon}^v t_z)\, (\mathbf{C}^v)^{T}  \notag \\
 &= |\omega_z|^{1/4}\, \exp(-t_z\, \mathbf{Q}\, \mathbf{F})\, \mathbf{Q}
\end{align}
are used to transform the two-electron integrals to write the $J$ and $K$ contributions
\begin{align}
e_J^{(z)} & = 2\, \sum_{\gm \gn \gk \gl} 
  (\underline{\gm} \bar{\gn} | \gk \gl)^{(z)}
  (\gm \gn | \underline{\gk} \bar{\gl})^{(z)} \text{,} \\
e_K^{(z)} & = \phantom{2}\, \sum_{\gm \gn \gk \gl} 
  (\underline{\gm} \bar{\gn} | \gk \gl)^{(z)}
  (\gm \bar{\gl} | \underline{\gk} \gn)^{(z)} 
\end{align}
in terms of half-transformed integrals (HTI)
\begin{align} 
(\underline{\gm} \bar{\gn} | \gk \gl)^{(z)} &=
  \sum_{\gm'} \po_{\gm'\gm}^{(z)} \left( \sum_{\gn'} \pv_{\gn\gn'}^{(z)}  (\gm' \gn' | \gk \gl) \right) \text{,} \label{htinr1} \\
(\gm \bar{\gl} | \underline{\gk} \gn)^{(z)} &=
  \sum_{\gk'} \po_{\gk'\gk}^{(z)} \left( \sum_{\gl'} \pv_{\gl\gl'}^{(z)}  (\gm \gl' | \gk' \gn) \right)  
\text{.} \label{htinr2} 
\end{align}
In Eqs.\ \eqref{psdeno_nr} -- \eqref{psdenv_nr}, 
$\mathbf{F}$ is the Fock matrix and $\mathbf{\boldsymbol\varepsilon}$ the orbital-energy vector;
$o$ and $v$ indicate the set of 
all active occupied and virtual orbitals, respectively.
In Eqs.\ \eqref{htinr1} -- \eqref{htinr2} and in the following 
$\gm$, $\gn$, $\gk$, and $\gl$ denote scalar AO basis functions.
Since the two density matrices $\mathbf{P}$ and $\mathbf{Q}$
are related through the constant AO overlap matrix $\mathbf{S}$ by
\begin{align}
& \mathbf{P} = \mathbf{C}^o \, (\mathbf{C}^o)^{T} \text{,} \\
& \mathbf{P} + \mathbf{Q} = \mathbf{S}^{-1}
\text{,}
\end{align}
the NR MP2 energy in Eq.\ \eqref{emp2lap} is a functional of the 
HF density matrix $\mathbf{P}$.\cite{Surjan2005}
To simplify the formulae, the index $z$ that denotes the 
quadrature point is omitted in the following when appropriate.

\subsection{LT AO-based MP2 for 2C Hamiltonians}
We can repeat this procedure for relativistic spinors in which we use the
Kramers' symmetry of the MO coefficients to reduce the number of operations.
If spinors are generated in a KR algorithm, the full spinor set
is subdivided in two sets of spinors with related coefficients (Eq. \eqref{LCAO}).  
This then leads to the classification of 16 different sub-blocks
of two-electron integrals $(IA|JB)$:
\begin{align} \label{intblock}
& (ia|jb)        \, (\ki\ka|jb)     \, (\ki a|\kj b)   \, (\ki a|j\kb)     \notag \\[2ex]
& (ia|j\kb)           \, (ia|\kj b)       \, (i\ka|jb)        \, (\ki a|jb)       \notag \\[2ex]
& (\ki\ka|\kj\kb)   \, (ia|\kj\kb)     \,  (i\ka|j\kb)    \, (i\ka|\kj b)  \notag \\[2ex]
& (\ki\ka|\kj b)     \, (\ki\ka|j \kb) \, (\ki a|\kj\kb) \, (i\ka|\kj\kb)  \text{,}
\end{align}
for which it is easy to show that the last 8 integral blocks are the complex conjugate
of the first 8. Furthermore, the integrals on the second and the fourth line are zero if
the system under consideration has (at least) a two-fold element of symmetry (mirror plane or
rotation axis).
The Coulomb energy
\begin{align} \label{ecoul1}
E_J = \frac{1}{2} \sum_{JIBA} \frac{1}{\Delta_{AIBJ}} &
   (BJ|AI) (IA|JB)  \\
=  \sum_{jiba} \frac{1}{\Delta_{aibj}} \Big(
  &   (ai|bj)(ia|jb)           +   (\ka\ki|bj)(\ki\ka|jb) \notag \\
+ & 2 (a\ki|bj)(\ki a|jb)      + 2 (\ka i|bj)(i\ka|jb)    \notag \\
+ &   (a\ki|b\kj)(\ki a|\kj b) +   (a\ki|\kb j)(\ki a|j\kb) \Big)
\end{align}
comprises 6 contributions that are real even without exploiting Kramers symmetry 
because one of the integrals in each of the products in Eq. \eqref{ecoul1} 
is the complex conjugate of the one it is multiplied with.
In the KR formalism, the exchange energy
\begin{align} \label{eexch1}
E_K 
=  \sum_{jiba} \frac{1}{\Delta_{aibj}} \Big(
  &   \re{(bi|aj)(ia|jb)}           +   \re{(b\ki|\ka j)(\ki\ka|jb)} \notag \\
+ & 2 \re{(b\ki|aj)(\ki a|jb)}      + 2 \re{(bi|\ka j)  (i\ka|jb)}   \notag  \\
+ &   \re{(b\ki|a\kj)(\ki a|\kj b)} +   \re{(\kb \ki|aj)(\ki a|j\kb)} \Big)
\end{align}
comprises 6 real contributions as the integral products in \eqref{eexch1}
appear in complex conjugate pairs with imaginary parts canceling each other.
The spinor-based integral products in Eqs.\ \eqref{ecoul1} and \eqref{eexch1} 
can be factorized in terms of two-electron integrals
and density matrices as for the NR case.
Here, we only show the working equations in their most condensed 
formulation of 2C AO-based MP2.
For a complete derivation we refer to the supporting information.\cite{si}

Before proceeding, we introduce quaternion algebra\cite{Saue1997,*Saue1999} for the KR formalism as
it is the most compact notation and results in working equations
that feature the least number of computational operations.
The quaternion spinor coefficients are given by
\begin{align}\label{qspinor}
{^q}\mathbf{C}
&= \mathbf{C}^{\alpha} - (\mathbf{C}^{\beta})^*\, j                  \notag \\
&= \re{\mathbf{C}^{\alpha}}    + \im{\mathbf{C}^{\alpha}}\ i
 - \re{\mathbf{C}^{\beta}}\ j  + \im{\mathbf{C}^{\beta}} \ k         \notag \\
&= \mathbf{C}^0 + \mathbf{C}^1\ i + \mathbf{C}^2\ j+ \mathbf{C}^3\ k
\end{align}
and from Eq.\ \eqref{qspinor} the quaternion density matrices
${^q}\mathbf{P}$ and ${^q}\mathbf{Q}$ can be constructed by
\begin{align}
& {^q}\mathbf{P} = {^q}\mathbf{C}^o \, ({^q}\mathbf{C}^o)^{\dag} 
= \mathbf{P}^0 + \mathbf{P}^1\ i + \mathbf{P}^2\ j+ \mathbf{P}^3\ k \text{,} \label{deno} \\
& {^q}\mathbf{P}+ {^q}\mathbf{Q} = \mathbf{S}^{-1}   \label{denv} 
\text{,}
\end{align}
where  $\mathbf{P}^0$,$\mathbf{Q}^0$ and 
$\mathbf{P}^{q_1}$,$\mathbf{Q}^{q_1}$ with $q_1 = 1,\ldots,3$
are real symmetric and anti-symmetric matrices, respectively.

As for the NR Hamiltonian,\cite{Surjan2005} the MP2 energy is a functional
of the quaternion HF density matrix ${^q}\mathbf{P}$, which facilitates a 
formulation and implementation without spinors purely in the AO basis. 
For relativistic 2C Hamiltonians, the Coulomb-
\begin{align} \label{ecoul}
 e_J^{(z)} = 2\, \sum_{\gm\gn\gk\gl} \sum_{\gm'\gn'\gk'\gl'}  \Big(
 & \po^0_{\gm' \gm} \pv^0_{\gn \gn'} - \po^1_{\gm' \gm} \pv^1_{\gn \gn'} 
 - \po^2_{\gm' \gm} \pv^2_{\gn \gn'} - \po^3_{\gm' \gm} \pv^3_{\gn \gn'} \Big) \notag \\
 \Big(
 & \po^0_{\gk' \gk} \pv^0_{\gl \gl'} - \po^1_{\gk' \gk} \pv^1_{\gl \gl'}
 - \po^2_{\gk' \gk} \pv^2_{\gl \gl'} - \po^3_{\gk' \gk} \pv^3_{\gl \gl'} \Big) \notag \\
 & (\gm'\gn'|\gk'\gl') (\gm\gn|\gk\gl) \\
    =  2\, \sum_{\gm\gn\gk\gl} \sum_{\gm'\gn'\gk'\gl'}  \Big(
 &  \re{{^q}\mathbf{\po}_{\gm' \gm}\, {^q}\mathbf{\pv}_{\gn \gn'}}
    \re{{^q}\mathbf{\po}_{\gk' \gk}\, {^q}\mathbf{\pv}_{\gl \gl'}} \Big) 
  (\gm'\gn'|\gk'\gl') (\gm\gn|\gk\gl)  \\
  = 2\, \sum_{\gm\gn\gk\gl} 
 & \re{ ({^q}\underline{\boldsymbol\gm} {^q}\bar{\boldsymbol\gn} | \gk \gl) }
   \re{    (\gm\gn | {^q}\underline{\boldsymbol\gk} {^q}\bar{\boldsymbol\gl}) }
\end{align}
and exchange contributions
\begin{align}  \label{eexch}
 e_K^{(z)} = \sum_{\gm\gn\gk\gl} \sum_{\gm'\gn'\gk'\gl'} \Big[ 
 \Big(
 & \po^0_{\gm' \gm} \pv^0_{\gl \gn'} - \po^1_{\gm' \gm} \pv^1_{\gl \gn'} 
 - \po^2_{\gm' \gm} \pv^2_{\gl \gn'} - \po^3_{\gm' \gm} \pv^3_{\gl \gn'} \Big)        \notag \\
 \Big(
 & \po^0_{\gk' \gk} \pv^0_{\gn \gl'} - \po^1_{\gk' \gk} \pv^1_{\gn \gl'}
 - \po^2_{\gk' \gk} \pv^2_{\gn \gl'} - \po^3_{\gk' \gk} \pv^3_{\gn \gl'} \Big)        \notag \\
 - \Big(
 &  \po^0_{\gm' \gm} \pv^1_{\gl \gn'} + \po^1_{\gm' \gm} \pv^0_{\gl \gn'} 
  - \po^2_{\gm' \gm} \pv^3_{\gl \gn'} + \po^3_{\gm' \gm} \pv^2_{\gl \gn'} \Big)       \notag \\
 \Big(
 &  \po^0_{\gk' \gk} \pv^1_{\gn \gl'} + \po^1_{\gk' \gk} \pv^0_{\gn \gl'}
  - \po^2_{\gk' \gk} \pv^3_{\gn \gl'} + \po^3_{\gk' \gk} \pv^2_{\gn \gl'} \Big)       \notag \\
 - \Big(
 &  \po^0_{\gm' \gm} \pv^2_{\gl \gn'} + \po^1_{\gm' \gm} \pv^3_{\gl \gn'} 
  + \po^2_{\gm' \gm} \pv^0_{\gl \gn'} - \po^3_{\gm' \gm} \pv^1_{\gl \gn'} \Big)       \notag \\
 \Big(
 &  \po^0_{\gk' \gk} \pv^2_{\gn \gl'} + \po^1_{\gk' \gk} \pv^3_{\gn \gl'}
  + \po^2_{\gk' \gk} \pv^0_{\gn \gl'} - \po^3_{\gk' \gk} \pv^1_{\gn \gl'} \Big)       \notag \\
 - \Big(
 &  \po^0_{\gm' \gm} \pv^3_{\gl \gn'} - \po^1_{\gm' \gm} \pv^2_{\gl \gn'} 
  + \po^2_{\gm' \gm} \pv^1_{\gl \gn'} + \po^3_{\gm' \gm} \pv^0_{\gl \gn'} \Big)       \notag \\
 \Big(
 &  \po^0_{\gk' \gk} \pv^3_{\gn \gl'} - \po^1_{\gk' \gk} \pv^2_{\gn \gl'}
  + \po^2_{\gk' \gk} \pv^1_{\gn \gl'} + \po^3_{\gk' \gk} \pv^0_{\gn \gl'} \Big) \Big] \notag \\
 & (\gm'\gn'|\gk'\gl') (\gm\gn|\gk\gl) \\
 = \sum_{\gm\gn\gk\gl} \sum_{\gm'\gn'\gk'\gl'} 
 & \re{{^q}\mathbf{\po}_{\gm' \gm}\, {^q}\mathbf{\pv}_{\gl \gn'}\,
       {^q}\mathbf{\po}_{\gk' \gk}\, {^q}\mathbf{\pv}_{\gn \gl'}   }                  
  (\gm'\gn'|\gk'\gl') (\gm\gn|\gk\gl) \\
 = \sum_{\gm\gn\gk\gl} 
 & \re{ ({^q}\underline{\boldsymbol\gm} {^q}\bar{\boldsymbol\gn} | \gk \gl)
      (\gm {^q}\bar{\boldsymbol\gl} | {^q}\underline{\boldsymbol\gk} \gn) }
\end{align}
are expressed in terms of two-electron integrals and occupied
\begin{align} \label{psdeno}
{^q}\mathbf{\po}^{(z)} 
 &= |\omega_z|^{1/4}\, {^q}\mathbf{C}^o \, \exp(+\mathbf{\boldsymbol\varepsilon}^o t_z)\, ({^q}\mathbf{C}^o)^{\dag}  \notag \\
 &= |\omega_z|^{1/4}\, \exp(+t_z\, {^q}\mathbf{P}\, {^q}\mathbf{F})\, {^q}\mathbf{P}
\end{align}
and virtual quaternion pseudo-density matrices
\begin{align} \label{psdenv}
{^q}\mathbf{\pv}^{(z)}  
 &= |\omega_z|^{1/4}\, {^q}\mathbf{C}^v\, \exp(-\mathbf{\boldsymbol\varepsilon}^v t_z)\, ({^q}\mathbf{C}^v)^{\dag}  \notag \\
 &= |\omega_z|^{1/4}\, \exp(-t_z\, {^q}\mathbf{Q}\, {^q}\mathbf{F})\, {^q}\mathbf{Q}
\text{.}
\end{align}

Apart from using quaternion rather than scalar Fock and density matrices,
the working equations to compute the SO 2C MP2 energy (Eqs.\ \eqref{ecoul} -- \eqref{eexch})
and the pseudo-density matrices (Eqs.\ \eqref{psdeno} -- \eqref{psdenv})
are identical to their NR counterparts.
The equations for computing the NR and spin-free (SF) MP2 energy
are identical because for the latter only the real part of the 
quaternion pseudo-densities will be non-zero. The quaternion formalism therefore
leads to an easy identification of SO contributions to the MP2 energy: these
are due to the imaginary parts of the pseudo-densities.

\subsection{Schwarz-type integral estimates}
In his seminal work on the LT AO-based MP2 H{\"a}ser
introduced integral estimates that are employed
for a Schwarz-type screening of transformed integrals.\cite{Haeser1993}
Due to the similarity of the working equations for NR and 2C MP2,
the original Schwarz-type screening can be extended easily
to discard contributions of AO quadruple to the Coulomb
\begin{align} \label{scrj}
& | (\underline{\gm}^{q_1} \bar{\gn}^{q_1} | \gk \gl) 
(\gm\gn | \underline{\gk}^{q_1} \bar{\gl}^{q_1})  |
 \le Z_{\gm \gn}^{q_1 q_1}\, Q_{\gk \gl} \, Q_{\gm \gn} \, Z_{\gk \gl}^{q_1 q_1}
\end{align}
and the exchange contribution 
\begin{align} \label{scrk}
& | (\underline{\gm}^{q_1} \bar{\gn}^{q_2} | \gk \gl) 
(\gm \bar{\gl}^{q_2} | \underline{\gk}^{q_1} \gn) | 
 \le Z_{\gm \gn}^{q_1 q_2}\, Q_{\gk \gl} \, Y_{\gm \gl}^{q_2} \, X_{\gk \gn}^{q_1}
\end{align}
to the 2C MP2 correlation energy.
The (quaternion pseudo-)Schwarz estimates in Eqs. \eqref{scrj} -- \eqref{scrk}
are given by:
\begin{align}
 \left(Q_{\gm \gn} \right)^2 &= 
   (\gm \gn| \gm \gn) \text{,} \label{estq} \\
\left(X_{\gm \gn}^{q_1} \right)^2 &=
 \po_{\gm' \gm}^{q_1} \po_{\gk' \gm}^{q_1} (\gm' \gn|\gk' \gn)  \text{,} \label{estx} \\
\left(Y_{\gm \gn}^{q_1} \right)^2 &=
 \pv_{\gn \gn'}^{q_1} \pv_{\gn \gl'}^{q_1} (\gm \gn'|\gm \gl')  \text{,} \label{esty}  \\
Z_{\gm \gn}^{q_1 q_2} &=
 \min( \sum_{\gn'} X_{\gm \gn'}^{q_1} |\pv_{\gn\gn'}^{q_2}| , 
       \sum_{\gm'} |\po_{\gm'\gm}^{q_1}| Y_{\gm' \gn}^{q_2} )
\label{estz} \text{.}
\end{align}
To keep the screening procedure computationally efficient
and also retain the rotational invariance of the MP2 energy,
the Coulomb (Eq. \eqref{scrj}) and exchange contributions
(Eq. \eqref{scrk}) should
be screened at the level of shells rather than individual 
AO basis functions.
Thus, for the estimates $\mathbf{Q}$, ${^q}\mathbf{X}$,
${^q}\mathbf{Y}$, and ${^{q, q}}\mathbf{Z}$ 
the maximum or Frobenius norm computed for all
basis functions that are associated with a shell pair
are employed for screening.

The Schwarz-type screening of Ref.\ \onlinecite{Haeser1993}
allows for an $\mathcal{O}(N^2)$ scaling with the system size $N$.
However, as it was shown for the NR LT AO-based MP2\cite{Doser2009,Maurer2013}
efficient linearly scaling implementations ($\mathcal{O}(N)$) 
are only in reach if one accounts for the physically 
correct decay behavior of two charge distributions separated by $R$.
For the exchange term $E_K$ fast exponential decay is expected.
The familiar $R^{-6}$ decay of $E_J$ at large $R$ is due to the
orthogonality of the occupied and virtual orbital space
which causes the zeroth-order moments of the multipole expansion 
of $1/R$ to vanish.\cite{Ayala1999,Lambrecht2005a} 
This holds for 2C LT AO-based MP2 as well and can be expressed
through an analogous orthogonality condition:
\begin{align} \label{zeroovlp}
 0 = {^q}\mathbf{ \po}^{(z)} \, \mathbf{S}\, {^q}\mathbf{ \pv}^{(z)} 
 \text{.}
\end{align}
Consequently, the efficient distance dependence adaptation of 
H{\"a}ser's integral estimates proposed by Ochsenfeld and 
his co-workers\cite{Maurer2013}
can be easily adapted to the 2C LT AO-based MP2 
implementation to achieve linear scaling eventually.

\section{Computational details}
The LT AO-based MP2 energy formulation
for relativistic 2C Hamiltonians was implemented in a
development version of DIRAC.
The two-electron integrals were computed with routines provided 
by the InteRest library.\cite{Repisky2013}
For all HF calculations the molecular mean field X2C Hamiltonian\cite{Sikkema2009}
was used.
For the preceding 4C HF calculations only the LL and LS type
two-electron integrals were calculated.\cite{Visscher1997}
Point-group symmetry was exploited at the HF level for
the HX molecules, only.
Nuclei were treated as Gaussian charge distributions.\cite{Visscher1997b}
For HX (X = F, Cl, Br, I, and At) 
we froze 2, 10, 18, 36, and 54~core electrons and 2, 6, 26, 40, and 96~anti-core 
electrons, respectively, and employed the all-electron double-zeta basis
set of Dyall.\cite{Dyallunpub,*Dyall2006,*Dyall2012}
Experimental equilibrium bond distances\cite{nistdiatom} were used for (HX, X=F,Cl,Br, and I);
0.9168~{\AA} (F), 1.2746~{\AA} (Cl), 1.4144~{\AA} (Br), and 1.6092~{\AA} (I).
The HAt bond distance 1.7075~{\AA} was taken from
an all-electron calculation of Peterson et al.\cite{Peterson2003}
The geometry of Hg-porphyrin was optimized with symmetry constraints
of the $D_{2h}$ point group by using the 
Turbomole package\cite{TM-V7.0} and is provided in the supporting information.\cite{si}
For the energy and gradient calculation we employed the dispersion-corrected\cite{Grimme2006,*Grimme2010} (D3)
PBE density functional\cite{Perdew1996} and the def2-TZVP basis set.
For the same calculation, an effective core potential\cite{Andrae1990} 
for Hg with 60 core electrons 
and multipole accelerated density fitting\cite{Eichkorn1995,*Eichkorn1995a,*Sierka2003} 
were used for reasons of performance.
For the AO-based MP2 calculation of Hg-porphyrin we used the SVP\cite{Schaefer1992} 
(H,C, and N) and all-electron double-zeta basis\cite{Dyall2012a}
sets (Hg) and froze 60~core and 24~anti-core electrons.
In case of Ba\tief{2} the all-electron double-zeta basis\cite{Dyall2012a} was employed
and the 72~core and 24~anti-core electrons were frozen.
The Laplace parameters $\omega_{z}$ and $t_{z}$
were obtained by the minimax approximation\cite{Takatsuka2008,*Helmich2016}
for a fixed number of quadrature points ($n_z=18$ for HX, $n_z=27$ for Ba\tief{2}, and
$n_z=1$ for Hg-porphyrin).
The errors in MP2 correlation energies for the diatomic molecules
caused by the numerical quadrature were always lower than $10^{-8}$~a.\ u. 
as was
verified by comparison with a reference (MO-based) implementation\cite{Laerdahl1997}.
For Hg-porphyrin only the quaternion Schwarz-type estimates were computed, the
MP2 correlation energy was not determined.

\section{Results and discussion}

\subsection{Comparison with spinor-based implementations}
Compared to the NR formulation, the LT AO-based 
formulation of MP2 energies for relativistic 2C Hamiltonians 
requires 16 times more operations for the computation of HTIs.
Contraction of HTIs to the MP2 energy has essentially the same scaling as NR for
the 2C Coulomb contributions $e_J^{(z)}$, and requires a 4 times larger 
effort for 2C exchange contributions $e_K^{(z)}$.

The ``SO factor'' of 16 in the time-determining
transformation step of our purely AO-based formulation
is similar to the factor of 16 found for the first half-transformation in the conventional
KR spinor-based approach\cite{Laerdahl1997,*Thyssen2008}. 
In both cases this arises from the need to use
a quaternion multiplication in the transformation of the second index, 
which takes $4\times 4=16$ terms
more operations than multiplication of real matrices. 
In the conventional formalism, also a second half-transformation
is required in which the quaternion unit of the 
second electron becomes active. 
This leads ultimately to
a factor of $16\times 4= 64$ if no symmetry can be used. 
In practice, one finds that for a spinor-based MP2 implementation
the most expensive step is often the initial quarter transformation,
which has a theoretical scaling factor of only 4. 
Furthermore, only part of the spinors are taken as
active in correlation calculation, thereby reducing the 
size of matrices as the index transformation proceeds.
Such reductions are particularly effective for heavy 
elements in which a number of core orbitals (and the associated
virtuals with high energies) can be frozen. 
For larger molecules, in which typically also many lighter elements
are present, the reduction in matrix size is smaller and the later 
steps of the index transformation become important. For 
such systems the AO-based formalism of the LT 
should therefore become competitive, also because integral screening
can be very effective with the highly localized densities of heavy elements.
It was shown for NR LT AO-based MP2
that integral screening leads to low-order\cite{Haeser1993,Ayala1999,Maurer2014}
or even linearly scaling implementations\cite{Doser2009,Maurer2013}
for medium- and large-sized systems. 
Thus, screening and the better SO scaling of the LT AO-based approach 
has the potential for early break-even points with spinor-based approaches.

\subsection{Effect of the nuclear charge on the energy contributions}
\begin{figure}[p]
 \centering
 \includegraphics[width=0.95\textwidth]{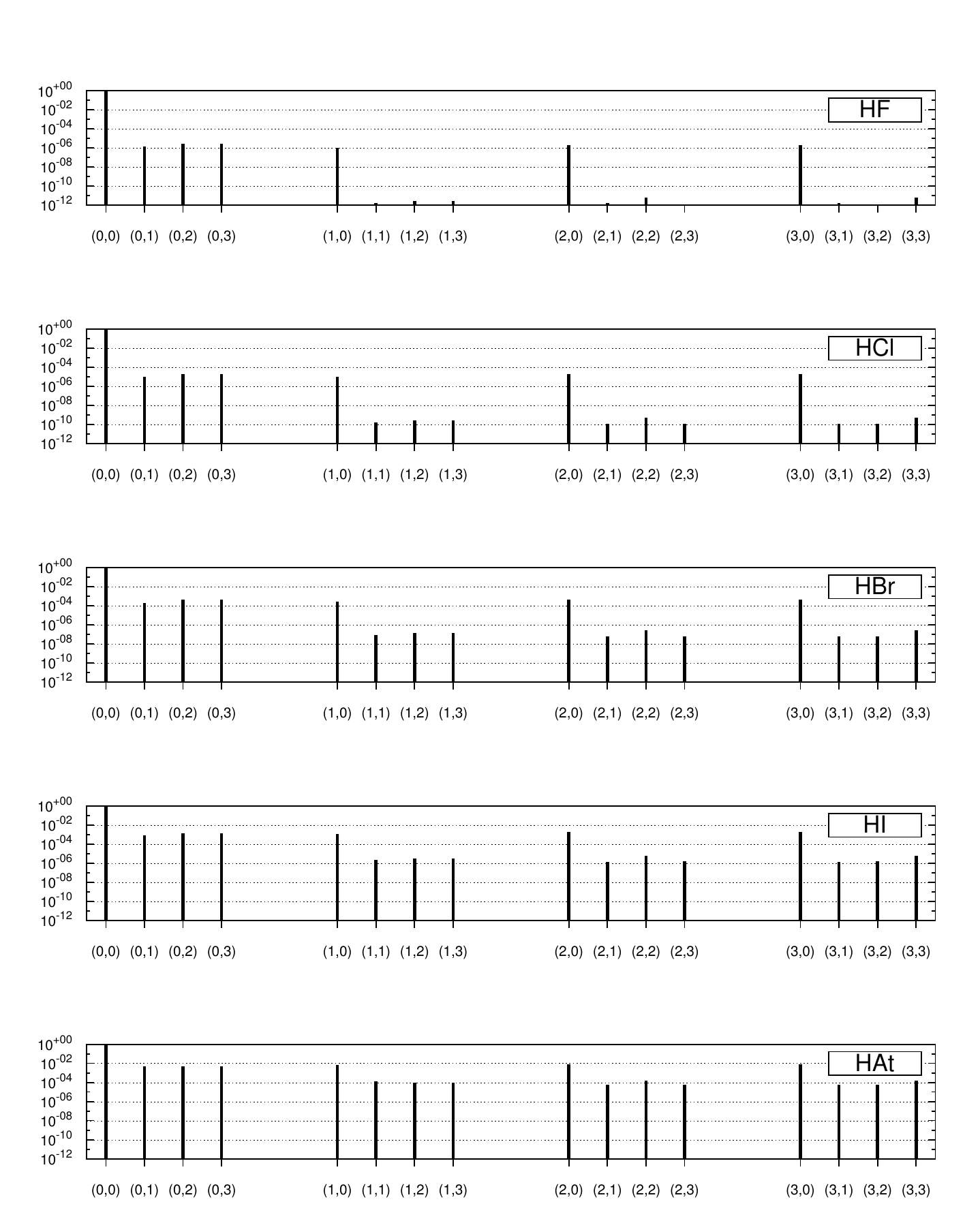}
 \caption{
$|({^q}\underline{\boldsymbol\gm} \, {^q}\bar{\boldsymbol\gn} | \gk \gl)|^2$
of the 16~HTIs summed over all quadrature points and normalized 
to the largest component $(0,0)$.
}
 \label{fig1}
\end{figure}

Our current implementation does not exploit any sparsity 
and has a hard $\mathcal{O}(N^5)$ scaling with the system size $N$.
To show the potential of screening on the computational work,
we analyze the 16~components of 
$|({^q}\underline{\boldsymbol\gm} \, {^q}\bar{\boldsymbol\gn} | \gk \gl)|^2$
from Eqs.\ \eqref{ecoul} -- \eqref{eexch}
for the halogen hydrides HX (X = F, Cl, Br, I, and At)
in Fig.\ \ref{fig1}.
Note that if the nuclei of a linear molecule are located
at one of the Cartesian coordinate axes, two of the three imaginary quaternion parts
will be identical for symmetry reasons.
For HX in Fig.\ \ref{fig1} as well as Ba\tief{2} in Figs.\ \ref{fig2} -- \ref{fig3}
we chose the z-axis as molecular axis; thus $q=2$ and $q=3$ are equivalent.
The real or SF parts of the quaternion pseudo-densities ${^q}\mathbf{\po}$ and ${^q}\mathbf{\pv}$ 
are much larger then the imaginary or SO parts, which allows for the following grouping 
of HTIs according to the 16~different combinations termed $(q_1,q_2)$
of ${^q}\mathbf{\po}$ and ${^q}\mathbf{\pv}$:
(I) only SF densities;
(II) one SF and one SO density;
(III) two SO densities.
As can be seen from Fig.\ \ref{fig1}, 
the difference is most distinct for the lightest hydride X=F where
(I) and (II) differ roughly by a factor of $10^{-6}$ 
while (I) and (III) differ by $10^{-12}$.
For such a light molecule, the energy contributions from HTIs that
contain only SO densities
are neglected without loss of accuracy if screening is exploited.
Since the mixed SF-SO density contributions are small as well,
the SF formalism suffices to describe the
2C MP2 energy sufficiently accurate for such a light molecule 
--- the relative deviation to SO 2C MP2 is $2.8\times 10^{-6}$ only.
The difference between (I) and (II)
HTIs and (II) and (III) HTIs decreases by one and two orders of magnitude,
respectively, each time we compare with a heavier homologue.
For the heaviest halogen hydride HAt the SF-only
HTIs and the mixed SF-SO HTIs differ by a factor of $10^{-2}$
while the former and the SO-only HTIs differ roughly by $10^{-4}$.
For HAt the relative deviation between SF and 2C MP2 is
much larger ($2.5\times 10^{-3}$) than for X=F.
Thus, the SF approximation shows only limited accuracy for this small molecule that
is dominated by the heavy element. For larger molecules in which one or more heavy
atoms are present a complete neglect of SO coupling
may likewise lead to significant errors. However, as all HTIs are expressed in the AO basis, 
it is still possible to obtain considerable savings by only neglecting SO contributions for
pseudo density matrix elements of the light elements. Such savings are not possible in
conventional MO-based electron correlation treatments.

\subsection{Distance dependence of the energy contributions and estimates}
\begin{figure}[p]
 \centering
 \includegraphics[height=0.75\textheight]{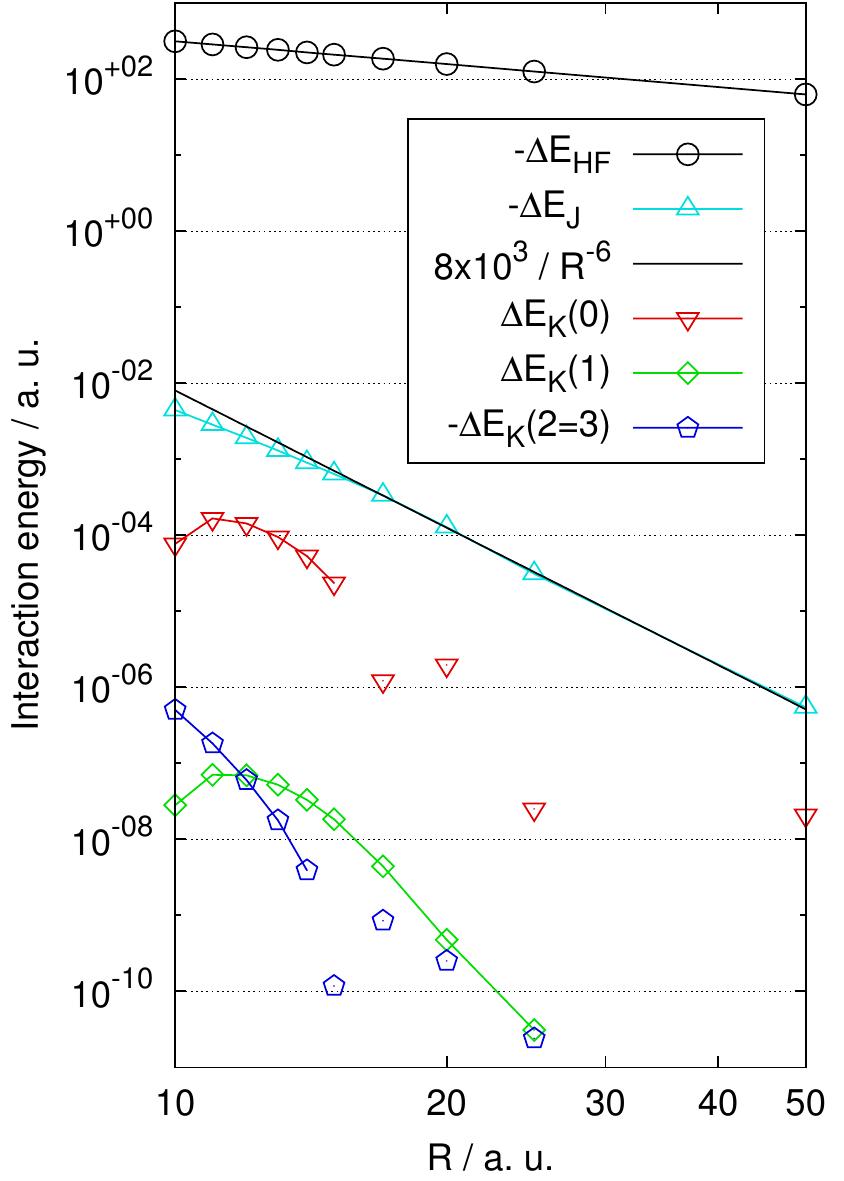}
 \caption{
$E_J$, $E_K$, and electronic HF contributions
to the SO 2C MP2 interaction energy at different inter-atomic distances $R$
of the barium dimer.
$\Delta E_K$ points not connected by a line have an opposite sign.
}
 \label{fig2}
\end{figure}

\begin{figure}[p]
 \centering
 \includegraphics[width=0.95\textwidth]{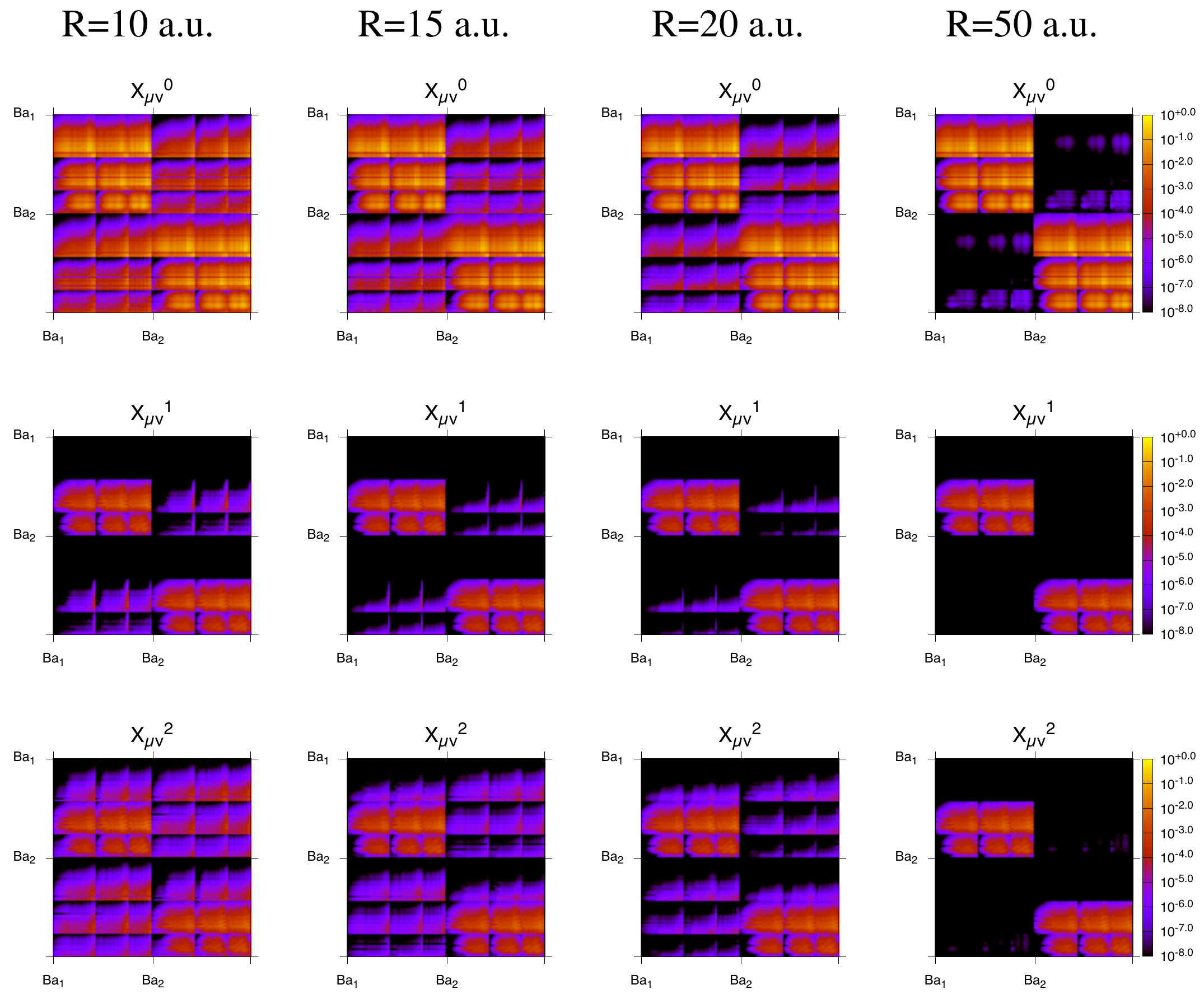}
 \caption{
Individual elements of the estimate ${^q}\mathbf{X}$ 
at different inter-atomic distances $R$
of the barium dimer.}
 \label{fig3}
\end{figure}

Besides the nuclear charge, the inter-electronic distance
has a substantial effect on screening and its utilization results
eventually in a linearly scaling relativistic MP2 implementation.\cite{Doser2009,Maurer2013}
First, this distance behavior is investigated by
means of the contributions to the interaction energy of Ba\tief{2} in Fig.\ \ref{fig2}.
$E_J$ shows the typical $R^{-6}$ decay of the dispersion
energy as explained in Sec. \ref{theory}, 
the 4 contributions to $E_K$ decay more rapidly than the Coulomb
contribution; but the expected exponential scaling 
is hard to observe as the values result from taking
differences between small numbers.
Amongst the 4 exchange contribution, the one that includes the SF-only
HTIs is clearly the largest while the other 3 contribute only little.
It is also clear that the decay of the already small SO contributions $E_K(q=1,2=3)$ is at least 
as fast as the SF $E_K(0)$ contribution.

Individual elements of the ${^q}\mathbf{X}$ estimates are shown in Fig.\ \ref{fig3}
for selected points of the Ba\tief{2} potential curve from Fig.\ \ref{fig2}.
Like in case of the halogen hydrides,
the SF part of the ${^q}\mathbf{X}$ estimates is largest
in magnitude, that is,
the largest element in $\mathbf{X}^0$ is always more than a factor of 30
larger than in $\mathbf{X}^{q_1}$ with $q_1 = 1,2=3$.
As the inter-atomic distance increases, the largest elements of
the two intra-atomic blocks remains almost constant whereas
the largest element of the two inter-atomic blocks decreases
and is at R=50~a.\ u.\ less than $1.4\times 10^{-6}$, $6.1\times 10^{-10}$,
and $1.4\times 10^{-7}$  for $\mathbf{X}^{q_1}$ with $q_1 = 0,1,2=3$, respectively.
Thus, individual elements of the SF and SO pseudo-Schwarz estimates, which are used to
screen contributions to the 2C MP2 energy, become smaller in magnitude
as the inter-atomic distance increases.
If contributions to the 2C MP2 energy are screened by means of SF and SO
pseudo-Schwarz estimates in Eqs.\ \eqref{estx} -- \eqref{estz}, 
reduced- or even linear-scaling implementations can be designed
as for NR AO-based MP2\cite{Doser2009,Maurer2013,Maurer2014}.

Like for HTIs in Fig.\ \ref{fig1} and for the contributions to the exchange interaction
energy in Fig.\ \ref{fig2} the SO part of the estimates $\mathbf{X}^{q_1}$ with $q_1 = 1,\ldots,3$
is much smaller than the SF part, too.
This will result in a much more efficient computation of the SO, rather than the SF,
energy contributions once screening is exploited.
Therefore, in practice, it is expected that the inherent ``SO factor'' of 4 and 16
for the integral contraction and transformation, respectively, 
is reduced significantly as there are relatively more SO contributions that fall
below threshold at larger distances than SF contributions.

\subsection{
Combined effect of the nuclear charge and 
the inter-electronic distance on screening
}
\begin{figure}[p]
 \centering
 \includegraphics[width=0.95\textwidth]{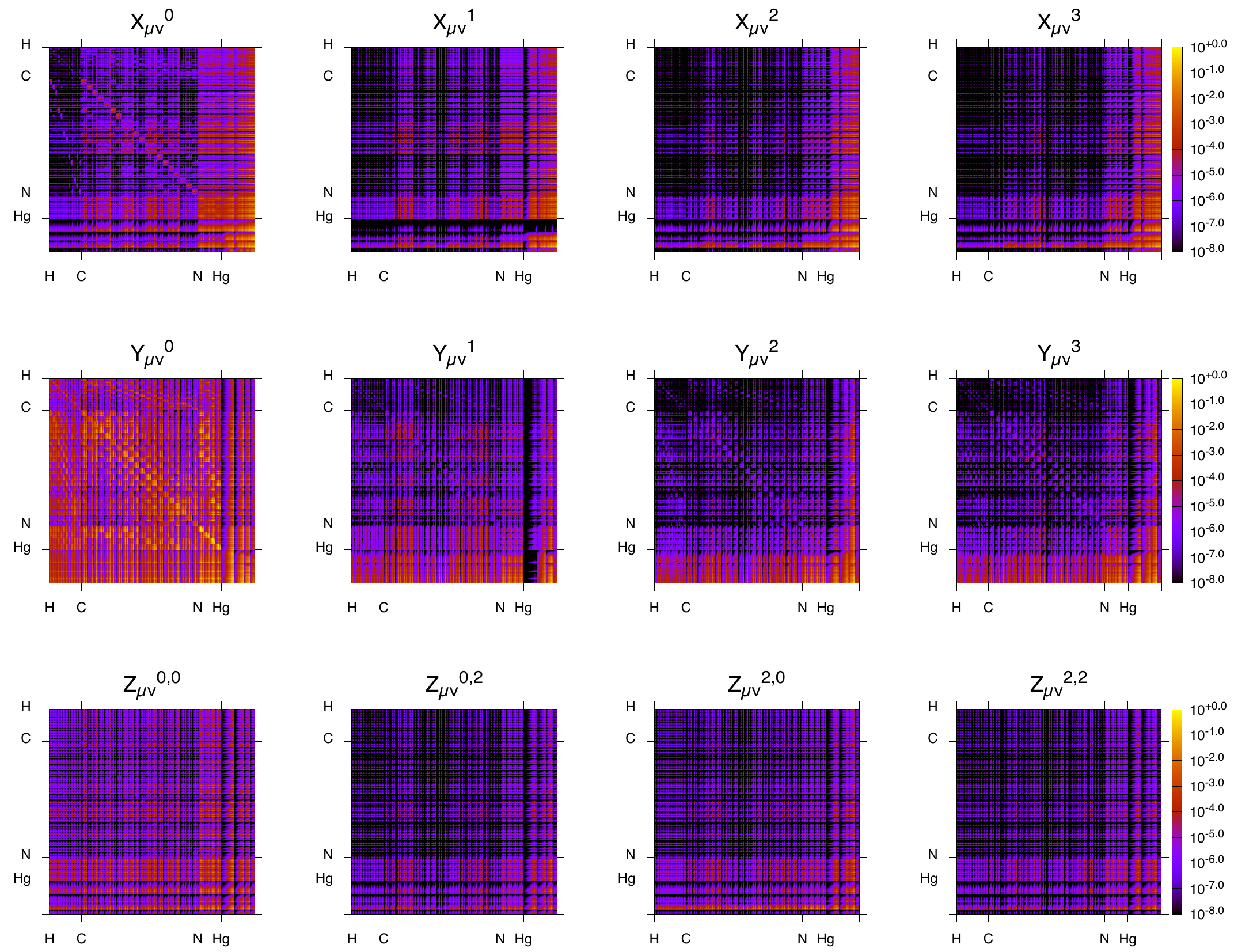}
 \caption{
Individual elements of the estimates ${^q}\mathbf{X}$ 
and ${^q}\mathbf{Y}$ and some selected  ${^{q,q}}\mathbf{Z}$
for Hg-porphyrin.
}
 \label{fig4}
\end{figure}

To illustrate the joint effects of the nuclear charge 
from different elements and the inter-electronic distances,
we calculated the estimates ${^q}\mathbf{X}$, ${^q}\mathbf{Y}$, 
and some of ${^{q,q}}\mathbf{Z}$ for Hg-porphyrin as depicted in Fig.\ \ref{fig4}.
Note that as for the linear molecules HX and Ba\tief{2} 
two of the three imaginary quaternion parts are identical ($q=2$ and $q=3$)
because the x- and y-axis of the molecular coordinate system
coincide with the $C_2$ symmetry elements of the $D_{2h}$ point group.

Since molecules that are mainly composed of a very few heavy metal elements
surrounded by organic ligands are highly relevant for synthetic and 
bio- organic and inorganic chemistry,
Hg-porphyrin is an ideal example to investigate. 
As expected, the shell pairs of light elements in porphyrin make only significant
contributions to the SF part of the estimates ($q=0$).
Consequently, the light elements will mainly contribute to SF part of the 2C MP2 energy
while Hg dominate the SO part of the 2C MP2 energy.

The largest element in any of the estimates is always one of the Hg-Hg shell pairs,
except for $\mathbf{Y}^0$ where the maximum value is one of the N-N shell pairs.
The H-H shell pairs contribute the least to the estimates,
especially for the SO type estimates.
Furthermore, it can be observed that
the closer the light atoms are to Hg the larger are their estimates.
For example, consider the C-C blocks, 
where the 8 C atoms that are neighbors of the N atoms 
have larger estimates than the remaining 12 C atoms located
at the outside and bridges of the porphyrin ring.
Since the estimates are employed to screen contributions of shell
quadruple to the MP2 correlation energy, we can conclude that
the largest contributions, both SF and SO, to the correlation energy 
originate from the Hg and N atoms and the least from the H atoms.
A more detailed analysis and also quantification of the number of 
screened shell quadruples is beyond the scope of the
present work and can be done once the distance-dependent extension
of the pseudo-Schwarz estimates is available.

\section{Conclusions}
We presented a formulation that is ready for combining
linear-scaling techniques for wave-function
methods with relativistic methods
that solve the Dirac equation explicitly.
Working equations are given for the LT AO-based 
MP2 for relativistic 2C Hamiltonians in the KR
formalism and compared to their NR counterparts.
By an analysis of the norm of HTIs
it is shown that for light molecules contributions
that arise from taking SO coupling into account
can be easily neglected by screening.
For molecules with heavy atoms those SO contributions are important and
will not be completely neglected when screening is exploited.
Furthermore, we have shown that the relativistic 2C Coulomb and exchange contributions
to the correlation energy
feature the same decay properties with an increasing separation
of charge distributions as their NR counterparts.
An adaptation of distance-dependent integral estimates developed
for NR MP2 to quaternion-based HTIs will therefore result in
a linearly scaling 2C MP2 implementation with early break-even points
compared to conventional spinor-based implementations.
Schwarz-type screening estimates can be adapted easily 
to the 2C LT AO-based MP2 formalism.
Both SF- and SO-type estimates become negligible for large inter-electronic
distances.
For light elements only the SF-type estimates show sizable
contributions,
which makes the 2C LT AO-based MP2 formalism attractive
for calculation on large metal-organic complexes
once a linearly scaling implementation is available.

\section{Acknowledgments}
Financial support from the German research foundation DFG
through grant number HE 7427/1-1 is gratefully acknowledged.
MR acknowledges support from the Research Council of Norway 
through a Centre of Excellence Grant (No. 179568/V30).
The authors thank Trond Saue for helpful discussions
on the quaternion formulation of the Dirac--Hartree--Fock method.

\bibliography{paper}

\newpage

\end{document}